\documentclass{aa}
\usepackage{graphicx}
\usepackage{txfonts}

\usepackage{natbib}
\usepackage{amsmath} 
\usepackage{amssymb}
\usepackage{mathtools}
\usepackage[dvipsnames]{xcolor}
\definecolor{CiteRed}{RGB}{110, 0, 0}
\usepackage{multirow}
\usepackage{array}
\usepackage{soul} 
\usepackage{makecell}
\usepackage{siunitx}
\usepackage{xcolor}     
\usepackage{orcidlink}  
\usepackage{ulem}
\usepackage{graphicx}
\usepackage{subcaption}
\usepackage{lipsum}
\usepackage{microtype}
\usepackage{verbatim}
\graphicspath{ {img/} {plot/} {fig/} }
\usepackage{hyperref}
\hypersetup{
    breaklinks,
    colorlinks,
    citecolor=CiteRed,  
    linkcolor=NavyBlue, 
    urlcolor=NavyBlue,  
    linktoc=page,
    pdftitle={Number of peaks in GRBs},
    pdfkeywords={GRBs},
    pdfauthor={Cristiano Guidorzi},
    pdfcreator={\LaTeX}
}

\begin{document}
\title{Distribution of number of peaks within a long gamma--ray burst}
\author{C.~Guidorzi\thanks{guidorzi@infn.it}\inst{\ref{inst1},\ref{inst2},\ref{inst3}}\orcidlink{0000-0001-6869-0835}
\and M.~Sartori\inst{\ref{inst1}}
\and R.~Maccary\inst{\ref{inst1}}\orcidlink{0000-0002-8799-2510}
\and A.~Tsvetkova\inst{\ref{inst4},\ref{inst3},\ref{inst5}}\orcidlink{0000-0003-0292-6221}
\and L.~Amati\inst{\ref{inst3}}\orcidlink{0000-0001-5355-7388}
\and L.~Bazzanini\inst{\ref{inst1},\ref{inst3}}\orcidlink{0000-0003-0727-0137}
\and M.~Bulla\inst{\ref{inst1},\ref{inst2},\ref{inst6}}\orcidlink{0000-0002-8255-5127}
\and A.~E.~Camisasca\inst{\ref{inst1}}\orcidlink{0000-0002-4200-1947}
\and L.~Ferro\inst{\ref{inst1},\ref{inst3}}\orcidlink{0009-0006-1140-6913}
\and F.~Frontera\inst{\ref{inst1},\ref{inst3}}\orcidlink{0000-0003-2284-571X}
\and C.~K.~Li\inst{\ref{inst7}}\orcidlink{0000-0001-5798-4491}
\and S.~L.~Xiong\inst{\ref{inst7}}\orcidlink{0000-0002-4771-7653}
\and S.~N.~Zhang\inst{\ref{inst7},\ref{inst8}}\orcidlink{0000-0001-5586-1017}}

\institute{Department of Physics and Earth Science, University of Ferrara, Via Saragat 1, I-44122 Ferrara, Italy \label{inst1}
\and  INFN -- Sezione di Ferrara, Via Saragat 1, 44122 Ferrara, Italy\label{inst2}
\and INAF -- Osservatorio di Astrofisica e Scienza dello Spazio di Bologna, Via Piero Gobetti 101, 40129 Bologna, Italy \label{inst3}
\and Department of Physics, University of Cagliari, SP Monserrato-Sestu, km $0.7$, 09042 Monserrato, Italy \label{inst4}
\and Ioffe Institute, Politekhnicheskaya 26, 194021 St. Petersburg, Russia \label{inst5}
\and INAF, Osservatorio Astronomico d'Abruzzo, via Mentore Maggini snc, 64100 Teramo, Italy \label{inst6}
\and Key Laboratory of Particle Astrophysics, Institute of High Energy Physics, Chinese Academy of Sciences, Beijing 100049, People's Republic of China \label{inst7}
\and University of Chinese Academy of Sciences, Beijing 100049, People's Republic of China \label{inst8}
}
\abstract
{The variety and complexity of long duration gamma--ray burst (LGRB) light curves (LCs) encode a wealth of information on the way LGRB engines release their energy following the collapse of the progenitor massive star. Thus far, attempts to characterise GRB LCs focused on a number of properties, such as the minimum variability timescale, power density spectra (both ensemble average and individual), or through different definitions of variability. In parallel, a characterisation as a stochastic process was pursued by studying the distributions of waiting times, peak flux, fluence of individual peaks that can be identified within GRB time profiles. Yet, the question remains as to whether the diversity of GRB profiles can be described in terms of a common stochastic process.}
{Here we address this issue by extracting and modelling for the first time the distribution of the number of peaks within a GRB profile.}
{We analysed four different GRB catalogues: {\it CGRO}/BATSE, {\it Swift}/BAT, {\it BeppoSAX}/GRBM, and {\it Insight-HXMT}. The statistically significant peaks were identified by means of well tested and calibrated algorithm {\sc mepsa} and further selected by applying a set of thresholds on signal-to-noise ratio. We then extracted the corresponding distributions of number of peaks per GRB.}
{Among the different models considered (power-law, simple or stretched exponential) only a mixture of two exponentials turned out to model all the observed distributions, suggesting the existence of two distinct behaviours: (i) an average number of $2.1\pm 0.1$ peaks per GRB (``peak poor'') and accounting for about 80\% of the observed population of GRBs; (ii) an average number of $8.3\pm 1.0$ peaks per GRB (``peak rich'') and accounting for the remaining 20\% of the observed population.}
{We associate the class of peak-rich GRBs with the presence of sub-second variability, which instead appears to be absent among peak-poor GRBs. The two classes could result from two different regimes through which GRB inner engines release energy or through which energy is dissipated into gamma-rays.}

\maketitle

\section{Introduction}
\label{sec:intro}
Gamma--ray burst (GRB) prompt emission is observed in at least two kinds of explosive transients: (i) the merger of a compact object binary system \citep{Eichler89,Paczynski91,Narayan92,LIGO-Fermi17}, (ii) the core collapse of some kinds of massive stars also known as ``collapsar''  \citep{Woosley93,Paczynski98,MacFadyen99,Yoon05}. While the prompt emission duration remains the strongest hint on the nature of the progenitor, in a few cases it turned out to be deceitful. To avoid confusion, the two progenitor families (i) and (ii) are often referred to as Type-I and Type-II GRBs, respectively \citep{Zhang06_nat}. Hereafter, we will consider only the latter class.

Modelling of thousands of GRB prompt emission time-resolved spectra proved that the observed variety is likely the result of different radiative processes in different GRBs, given the variety of spectral components that are observed: the non-thermal Band function, an occasional quasi-thermal component, and even a broad-band power-law (see \citealt{KumarZhang15rev} for a review). These processes are possibly related to different dissipation mechanisms taking place at different distances from the progenitor and driven by key properties, such as the ejecta composition of the relativistic jet that is launched by the GRB inner engine.

In parallel, while numerous studies that focused on the temporal properties of GRB time profiles have contributed to characterise their variability, the large variety and highly erratic nature of GRB prompt emission remains mostly unexplained and apparently disconnected from the properties that are inferred from the afterglow modelling, apart from the energetics. Despite being affected by a large scatter, a positive correlation between peak luminosity and variability was found as soon as the number of bursts with measured redshift increased to a dozen or more \citep{Reichart01,Fenimore00}. Variability was defined as the net (that is, removed of the counting statistics noise contribution) variance of the light curve (LC) with respect to a smoothed version of the same. The correlation was later confirmed over larger data sets, along with its large scatter \citep{Guidorzi05b,Guidorzi05c,Guidorzi06b}, with a forthcoming analysis of {\it Swift} and {\it Fermi} GRBs (Guidorzi et al. in prep.). Peak luminosity was also found to correlate with so-called minimum variability timescale (MVT), defined as the shortest timescale over which an uncorrelated flux change is observed significantly in excess of statistical fluctuations (see \citealt{Camisasca23} and references therein). A possible interpretation explains more variable GRBs as the result of narrower jets with higher Lorentz factors and shorter spreading timescales than wider jets \citep{Kobayashi02}.

Most GRB LCs can be described as a sequence of peaks of different shapes, duration, and intensity with no clear rule. While different techniques have revealed that temporal power is distributed in general over a broad range of timescales (from $\sim 10$~ms to several $10$~s), two distinct components were also identified in a number of GRBs: slow ($\gtrsim 1$~s), and fast (down to few $10$~ms; \citealt{Vetere06,Gao12,Guidorzi16}). Since the first catalogues of the Burst And Transient Source Experiment (BATSE; \citealt{Paciesas99}) that flew aboard the Compton Gamma--Ray Observatory (CGRO; 1991--2000), distributions of properties of the peaks, seen as building blocks of the prompt emission, were studied in detail (e.g., see \citealt{RamirezRuiz01a,Quilligan02,NakarPiran02a}). Yet, despite the enormous progress in the knowledge of GRB progenitors and their environments, a precise characterisation of the stochastic process(es) that drive(s) GRB engines in releasing their energy as a function of time is missing. Hence, the degree of complexity exhibited by a given GRB LC, especially if it is seen as a point process given by the sequence of peaks with different intensities and durations, can hardly be translated into clues on the nature of its inner engine or on the nature of the powering mechanism that rules its evolution. 

In this respect, while the distribution of the number of peaks per GRB was obtained in past investigations (e.g., \citealt{Quilligan02,Guidorzi16}), to our knowledge no attempt to model and interpret it has been done to date. Following previous analogous definitions \citep{Li96,NakarPiran02a}, a ``peak'' is defined as a statistically significant increase of the count rate with respect to the neighbouring time bins, using the terminology introduced by \citet{Li96}. In this work we addressed this by applying a well tested peak identification code, {\sc mepsa} \citep{Guidorzi15a}, to four independent catalogues among the different past and present missions aimed at GRB studies \citep{Tsvetkova22}: {\it CGRO}/BATSE, {\it Swift}/BAT, {\it BeppoSAX}/GRBM, and {\it Insight-HXMT}. A forthcoming paper will report the still on-going analysis of {\it Fermi}/GBM GRBs, some preliminary results of which will be mentioned in what follows. We point out that our goal deliberately ignores the temporal structures of peaks, which then might differ significantly from one another, or how much they overlap in time, provided that they can be identified as distinct peaks. Section~\ref{sec:data} reports the data sets used, the LC extraction along with the peak identification. Section~\ref{sec:res} reports the results of the analysis, which are then discussed in Section~\ref{sec:disc_conc}.
  
\section{Data sets}
\label{sec:data}
We obtained the background-subtracted LC of GRBs with a bintime of 64~ms (except for {\it BeppoSAX}/GRBM for which we made use of both 1-s and $62.5$-ms bin times) and extracted a list of peaks detected with {\sc mepsa}, a code specifically devised and optimised to identify peaks in uniformly sampled and background-subtracted GRB LCs \citep{Guidorzi15a}, discarding candidates with a signal-to-noise ratio S/N $< 5$.

From the BATSE 4B catalogue~\citep{Paciesas99} we took the 64-ms time profiles of LGRBs ($T_{90}>2$~s) made available by the BATSE team\footnote{\url{https://heasarc.gsfc.nasa.gov/FTP/compton/data/batse/ascii_data/64ms/}}. Taken with the eight BATSE Large Area Detectors (LADs), these time profiles resulted from the concatenation of three standard BATSE types, DISCLA, PREB, and DISCSC. They are available in four energy channels, 25--55, 55--110, 110--320, and $> 320$~keV: we used the summed counts over the four channels. The background was interpolated with polynomials of up to fourth degree as prescribed by the BATSE team. These background-subtracted profiles were then processed with {\sc mepsa} and we selected the GRBs with at least one significant peak. We ended up with a sample of 1457 GRBs, which hereafter are referred to as the BATSE sample.

We started from all the GRBs detected by {\it Swift}/BAT in burst mode from January 2005 to July 2023 and rejected those with $T_{90}\le 2$~s as well as the long-lasting Type-I candidates, as they were classified by the {\it Swift} team: they include all the so-called short extended-emission GRBs \citep{Norris06}, in addition to peculiar events like 060614, 211211A \citep{Gehrels06,Yang22}. The information on the $T_{90}$ duration and on the LC characterisation was taken either from the BAT3 catalogue \citep{Lien16} when available, or from the {\it Swift}/BAT team circulars.
For each burst we extracted the mask-weighted, 64-ms bin time profile in the 15--150~keV passband following the standard procedure recommended by the BAT team\footnote{\url{https://swift.gsfc.nasa.gov/analysis/threads/bat\_threads.html}.} and applied {\sc mepsa}. Imposing the above mentioned S/N$>5$ threshold on the peak candidates, we selected all the Type-II GRB candidates with at least one peak and ended up with 1277 bursts. Hereafter, this will be referred to as the BAT sample.

The LCs available in the 40--700~keV passband for the GRBs from the {\it BeppoSAX}/GRBM catalogue \citep{Frontera09} come after two fashions: for the 2/3 that triggered the on-board logic, a $7.8125$-ms profile covering the first 106~s for each of the four independent units is available. For the remaining 1/3 only 1-s LCs are available. For the former we obtained $62.5$-ms profiles. 
The GRBs observed before November 1996 were ignored due to the presence of extra-Poissonian noise in the background rate of the GRBM, that was removed after raising the LLT energy threshold \citep{Frontera09}. We then discarded short GRBs ($T_{90}<2$~s, as reported by \citealt{Frontera09}) and ran {\sc mepsa} on the LC of the most illuminated detector (or on the sum of the two most illuminated ones, depending on which LC had the higher S/N) independently on both $62.5$-ms and 1-s profiles and merged the results, making sure to count once the peaks that triggered both timescales. The reason of this strategy is not to lose peaks that occurred after the first 106~s acquired with high temporal resolution. We applied the same $S/N>5$ threshold as for the other catalogues and ended up with 820 GRBs with at least one significant peak, which hereafter are referred to as the {\it BeppoSAX} sample.

From the first {\it Insight-HXMT}/HE GRB catalogue \citep{Song22_HXMTGRBcatalog} we discarded all the GRBs with $T_{90}<2$~s as well as those belonging to the iron sample, for which the electronics saturated due to excessively high count rates. For the remaining GRBs we extracted the 64-ms, dead-time corrected LCs obtained by summing the counts of all the 18 HE detector units, selecting only the CsI events (for which the HE works as an open-sky monitor) following the procedure described in \citet{Song22_HXMTGRBcatalog}. The total energy passband depends on the HE operation mode: either 80--800 or 200--3000 keV for the normal and GRB (or low-gain) mode, respectively. The background interpolation and subtraction was carried out as in \citet{Camisasca23}, to which the reader is referred. We finally ran {\sc mepsa} and after applying the same S/N$>5$ selection we ended up with 202 GRBs hosting at least one significant peak. Hereafter these GRBs are referred to as the {\it Insight-HXMT}\ sample.

\section{Results}
\label{sec:res}
For each data set we obtained the distribution of the number of peaks per GRB, expressed as number of GRBs having $n$ peaks, as a function of $n$ for all integer values ($n\ge 1$). This is clearly a discrete distribution. For each of the models considered below, we first determined the best fit parameters by maximising the corresponding likelihood function. We then estimated the parameters' marginalised distributions by sampling the posterior distribution using Markov Chain Monte Carlo (MCMC) sampling with {\sc emcee} \citep{Foreman13}: starting from the maximum likelihood solution, we run 32 walkers for 5000 steps and discard the ﬁrst 100 steps as burn-in. We tested the goodness of the best solution by grouping the expected number of GRBs so as to have at least 5 per class and carried out a $\chi^2$ test.

We first tried to model the distributions with a discrete simple exponential, which is mostly determined by the majority of GRBs, which have 1 or 2 peaks. However, all data sets showed an excess of GRBs with many ($\gtrsim 10$) peaks. Motivated by the possible presence of a hard tail, we then considered a simple power-law model, but the result was poor. A broken power-law did not provide satisfactory fits for all data sets either.

As a further attempt, we considered a stretched exponential (also known as the empirical Kohlrausch-Williams-Watts, or KWW, function), which is often encountered in several contexts in physics for modelling relaxation processes of systems that dissipate energy and are initially out of equilibrium (e.g. see \citealt{Lukichev19} and references therein). It can also result from the combination of numerous (ideally continuous) exponentially distributed processes \citep{Johnston06}. Moreover, a stretched exponential was found to describe the average temporal decay of peak-aligned and peak-normalised profiles of BATSE GRBs \citep{SternSvensson96}. As a result, only the distribution of the BATSE sample, which is the most numerous one, could not be modelled satisfactorily with a stretched exponential ($\chi^2$ test p-value of $10^{-6}$).

\begin{table*}
\caption{Best fit values and 90\% confidence intervals of the parameters of the mixture model of two exponentials (Eq.~\ref{eq:exp2}) applied to the different data sets.} 
\label{tab:fit_doublexp}
\centering 
\begin{tabular}{lrrcccccccc}
\hline\hline
Catalog  & $N^{\rm{(a)}}$ & $N_{p}^{\rm{(b)}}$ & $k$ & $n_1$ & $\langle n^{(1)}\rangle^{\rm{(c)}}$ & $n_2$  & $\langle n^{(2)}\rangle^{\rm{(c)}}$ & $\xi$ & $\bar{w}_2^{\rm{(d)}}$ & $\chi^2$ test\\
& & & & & & & & &   & p-value\\ 
\hline
BATSE      & 1457 & 8112 & $0.33_{-0.03}^{+0.02}$ & $2.89_{-0.26}^{+0.29}$ & $3.42_{-0.26}^{+0.29}$ & $13.4_{-2.1}^{+3.0}$ & $13.9_{-2.1}^{+3.0}$ & $0.049_{-0.018}^{+0.027}$ & $0.21_{-0.05}^{+0.06}$ & $0.164$ \\
BAT        & 1277 & 3817 & $0.78\pm 0.08$ & $1.47\pm 0.17$ & $ 2.03\pm 0.17$ & $ 6.4_{-1.2}^{+1.5}$ & $ 6.9_{-1.2}^{+1.5}$ & $0.043_{-0.017}^{+0.029}$ & $0.20_{-0.06}^{+0.08}$ & $0.845$ \\
BeppoSAX   &  820 & 2151 & $1.02_{-0.12}^{+0.13}$ & $1.29_{-0.14}^{+0.15}$ & $1.85_{-0.13}^{+0.15}$ & $ 7.6_{-1.9}^{+3.4}$ & $ 8.1_{-1.9}^{+3.4}$ & $0.019_{-0.011}^{+0.015}$ & $0.13_{-0.05}^{+0.06}$ & $0.440$ \\
HXMT       &  202 & 616 & $0.99_{-0.46}^{+0.71}$ & $0.85_{-0.23}^{+1.02}$ & $1.45_{-0.20}^{+0.96}$ & $ 4.3_{-1.2}^{+3.8}$ & $ 4.8_{-1.2}^{+3.8}$ & $0.107_{-0.076}^{+0.219}$ & $0.43_{-0.26}^{+0.23}$ & $0.600$ \\
All$^{\rm{(e)}}$   & 3756 & 14696 & $0.59\pm 0.03$ & $1.81_{-0.10}^{+0.11}$ & $2.36\pm 0.10$ & $ 9.6_{-0.9}^{+1.2}$ & $ 10.1_{-0.9}^{+1.2}$ & $0.037_{-0.008}^{+0.010}$ & $0.20\pm 0.03$ & $0.033$ \\
W. Aver.$^{\rm{(f)}}$ & -- & -- & -- & -- & $2.1\pm 0.1$ & -- & $8.3\pm 0.1$ & -- & $0.19\pm 0.03$ & --\\
\hline
\end{tabular}
\begin{list}{}{}
   \item [$^{\rm (a)}$] Number of GRBs.
   \item [$^{\rm (b)}$]  Total number of peaks.
   \item [$^{\rm (c)}$] Expected number of peaks per GRB of the corresponding exponential component, calculated as in Eq.~(\ref{eq:mean_exp}).
   \item [$^{\rm (d)}$] Defined in Eq.~(\ref{eq:waver}), it is the total fraction of GRBs contributed by the second exponential, characterised by a higher average number of peaks per GRB.
   \item [$^{\rm (e)}$] All data sets have been merged together and treated as one sample.
   \item [$^{\rm (f)}$] Weighted average value using the different estimates of the four catalogues.
\end{list}
\end{table*}

\begin{figure*}
    \centering
    \begin{tabular}{cc}
        \includegraphics[width=0.48\textwidth]{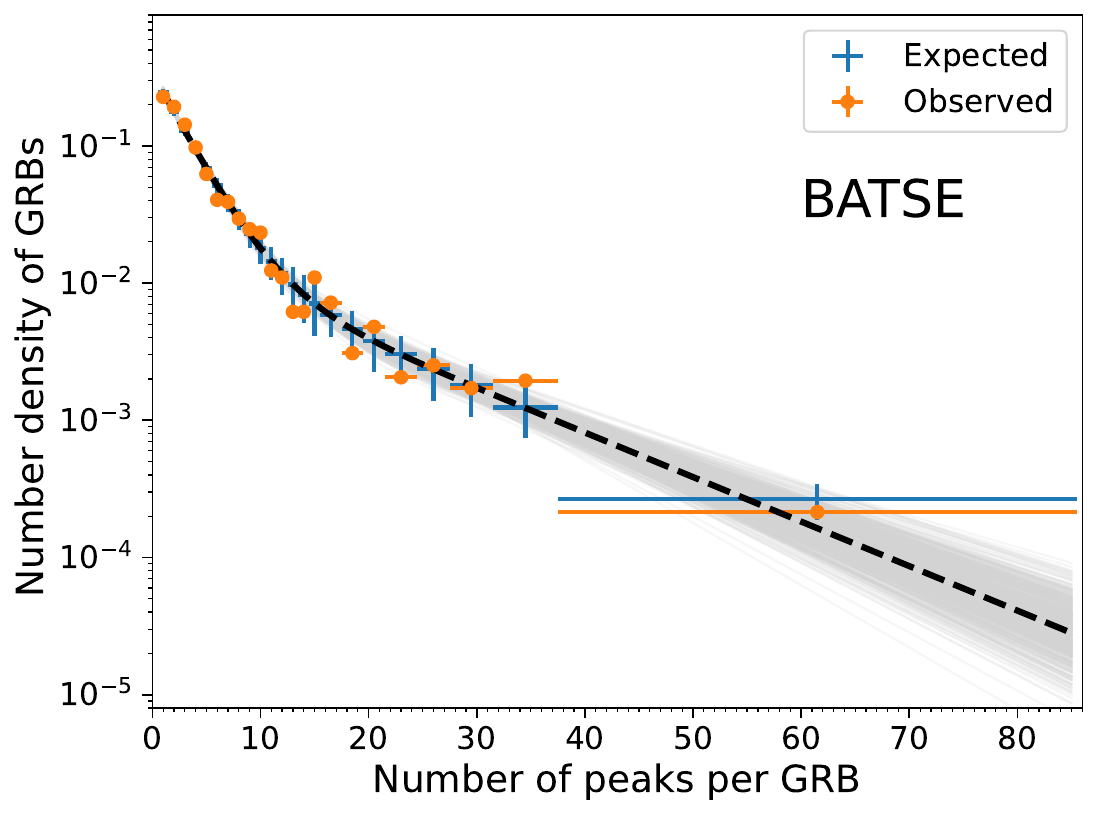}
        \includegraphics[width=0.48\textwidth]{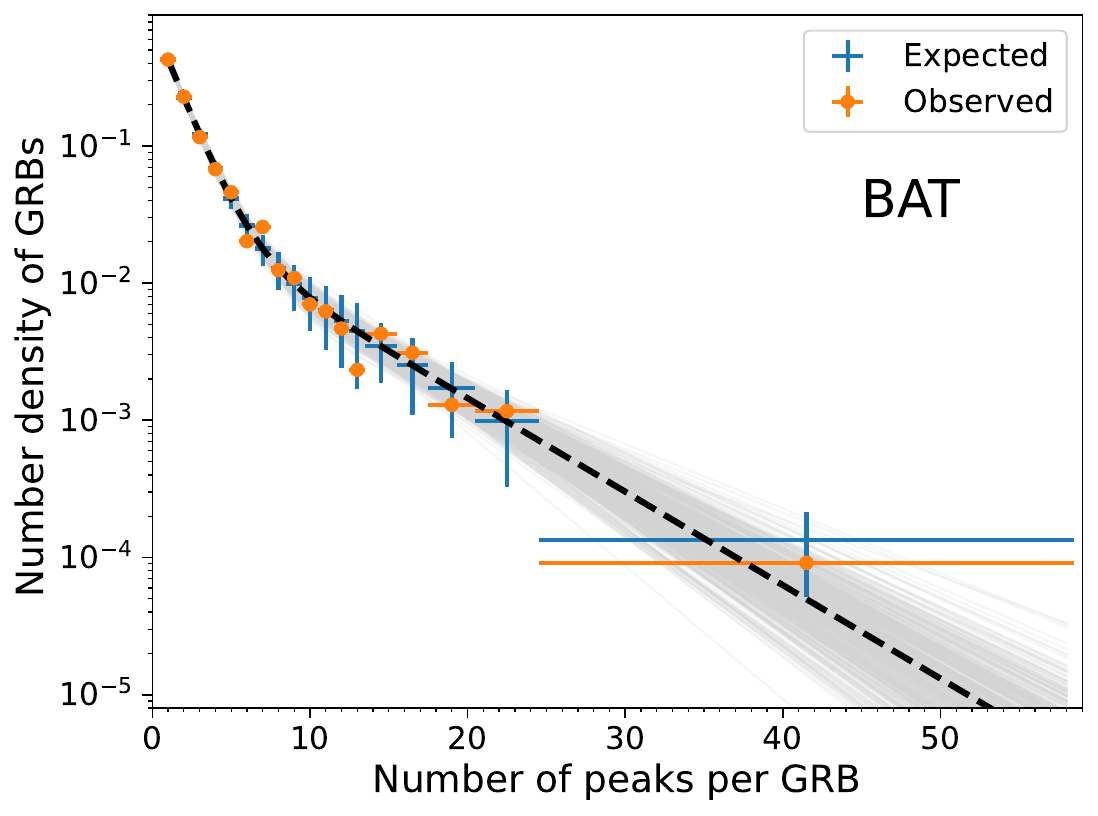} \\
        \includegraphics[width=0.48\textwidth]{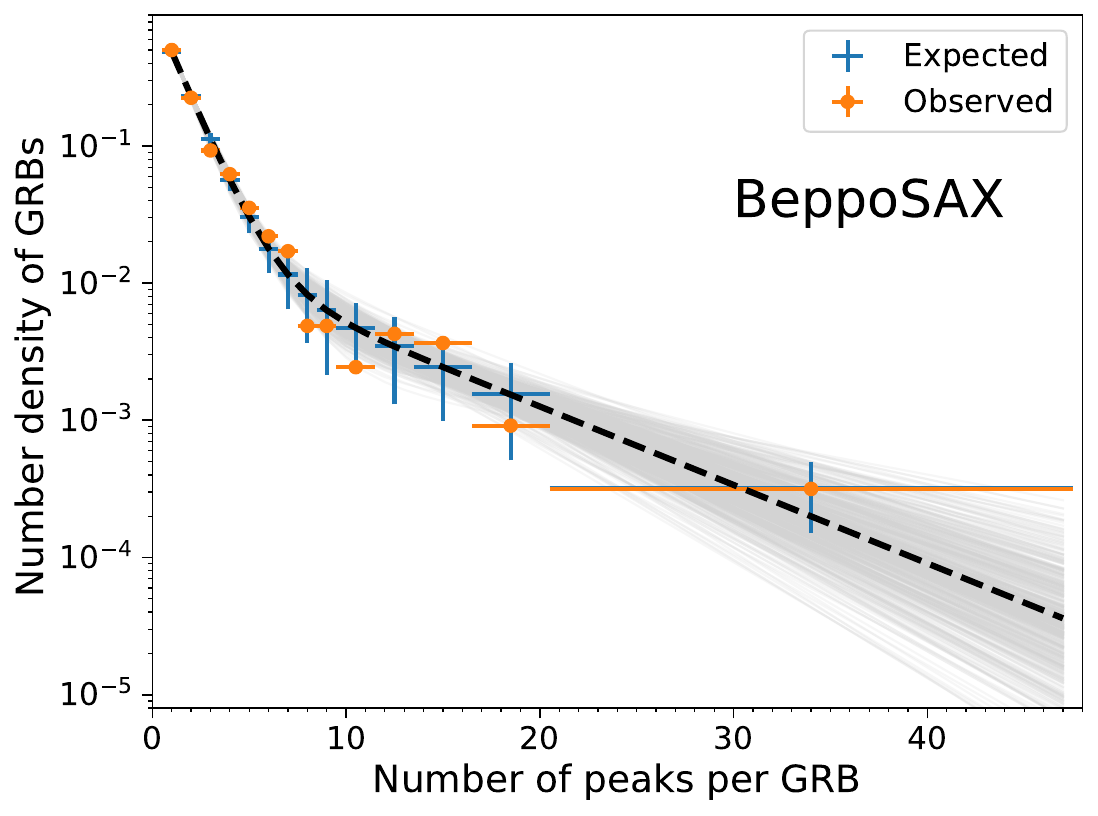}
        \includegraphics[width=0.48\textwidth]{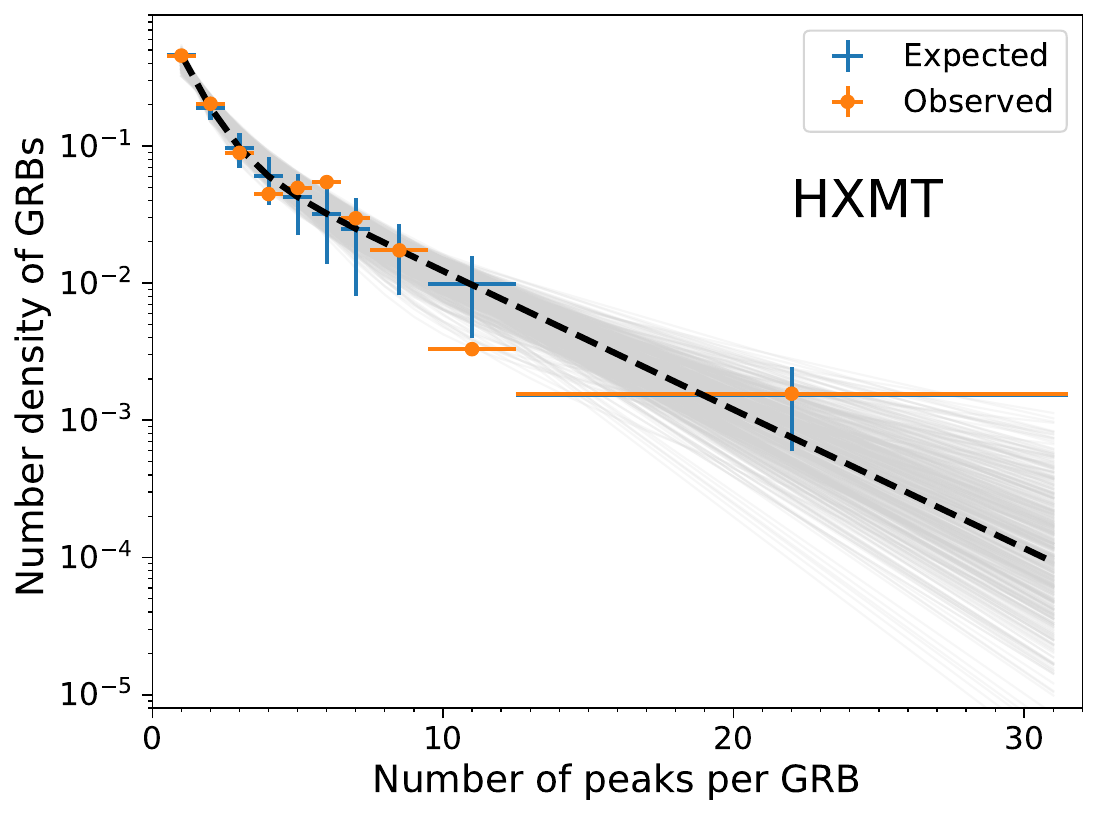}
    \end{tabular}
    \caption{Distributions of number of GRBs per number of peaks for different catalogues: {\it CGRO}/BATSE (top left), {\it Swift}/BAT (top right), {\it BeppoSAX}/GRBM (bottom left), {\it Insight-HXMT} (bottom right). Observed data and expected counts are shown in orange and blue, respectively. All histograms have been grouped so as to ensure a minimum number of counts per bin. Dashed lines show the best mixture model of two exponentials that fit each distribution. Grey models are the result of a random sampling of the posterior distribution of the parameters as determined via MCMC.}
    \label{fig:double_exp_individ}
\end{figure*}

We finally considered the superposition of two simple exponentials with a relative weight treated as a free parameter. Specifically, the fraction $f(n)$ of GRBs hosting $n$ peaks is described by Eq.~(\ref{eq:exp2}):
\begin{equation}
    f(n)\ =\ k\ ({e^{-n/n_1} \ +\ \xi \,e^{-n/n_2}})\;,
    \label{eq:exp2}
\end{equation}
where $k$ is a normalisation constant, such that $\sum_{n=1}^{+\infty} f(n) = 1$, $n_i$ ($i=1,2$) is the characteristic number of peaks per GRB of the $i$-th component ($n_1\le n_2$ by convention) and $\xi$ is a relative normalisation parameter, thus totalling 3 free parameters. 
Due to the discrete nature of the distribution, the corresponding expected number of peaks per GRB of the $i$-th exponential component, $\langle n^{(i)} \rangle$, is not simply $n_i$ as in the case of a continuous distribution, but the following:
\begin{equation}
\langle n^{(i)} \rangle\ =\ \frac{\sum_{n=1}^{+\infty} n e^{-n/n_i}}{\sum_{n=1}^{+\infty} e^{-n/n_i}}\ = \frac{1}{1-e^{-1/n_i}}\;.
\label{eq:mean_exp}
\end{equation}
Hereafter, Equation~(\ref{eq:exp2}) is referred to as the  mixture model of two exponentials (M2E).

The resulting distributions along with the best fit M2Es are shown in Figure~\ref{fig:double_exp_individ}, while Table~\ref{tab:fit_doublexp}
reports the best fit values of the model parameters and the corresponding p-value of the $\chi^2$ test. All of the four data sets can be satisfactorily modelled and, what is more, the parameters values are similar to each other, despite the different effective areas and energy passbands. We also performed a joint fit of all the four distributions with a common set of parameters in two ways: (i) by merging all GRBs and treating it as a single data set, thus dominated by the richest data sets; (ii) by keeping the four distributions separate and by assigning equal weights in the total likelihood. Only for (i) we obtained a marginally acceptable solution (p-value of 3\%, see Table~\ref{tab:fit_doublexp}).

We also calculated the fraction of GRBs $\bar{w}_2$ that are contributed by the second exponential component, according to each best fit solution: this is calculated simply as
\begin{equation}
    \bar{w}_2\ =\ k\ \xi \sum_{n=1}^{+\infty} e^{-n/n_2}\;.
    \label{eq:waver}
\end{equation}
Interestingly, all data sets have compatible values, with a weighted average of $0.19\pm 0.03$ (90\% confidence).
Analogous results are found for the expected numbers of peaks of either component: while the first component presents a range of values for $\langle n^{(1)}\rangle$ only marginally compatible with each other (weighted average of $2.1\pm0.1$, p-value of $6\times 10^{-6}$), the expected value for the second component, $\langle n^{(2)}\rangle$, is compatible among all data sets, with a weighted average of $8.3\pm 1.0$ (p-value of $0.06$). Hereafter, the first ($i=1$) and the second ($i=2$) components are referred to as the peak-poor and the peak-rich ones, respectively.
\begin{figure}
   \includegraphics[width=0.47\textwidth]{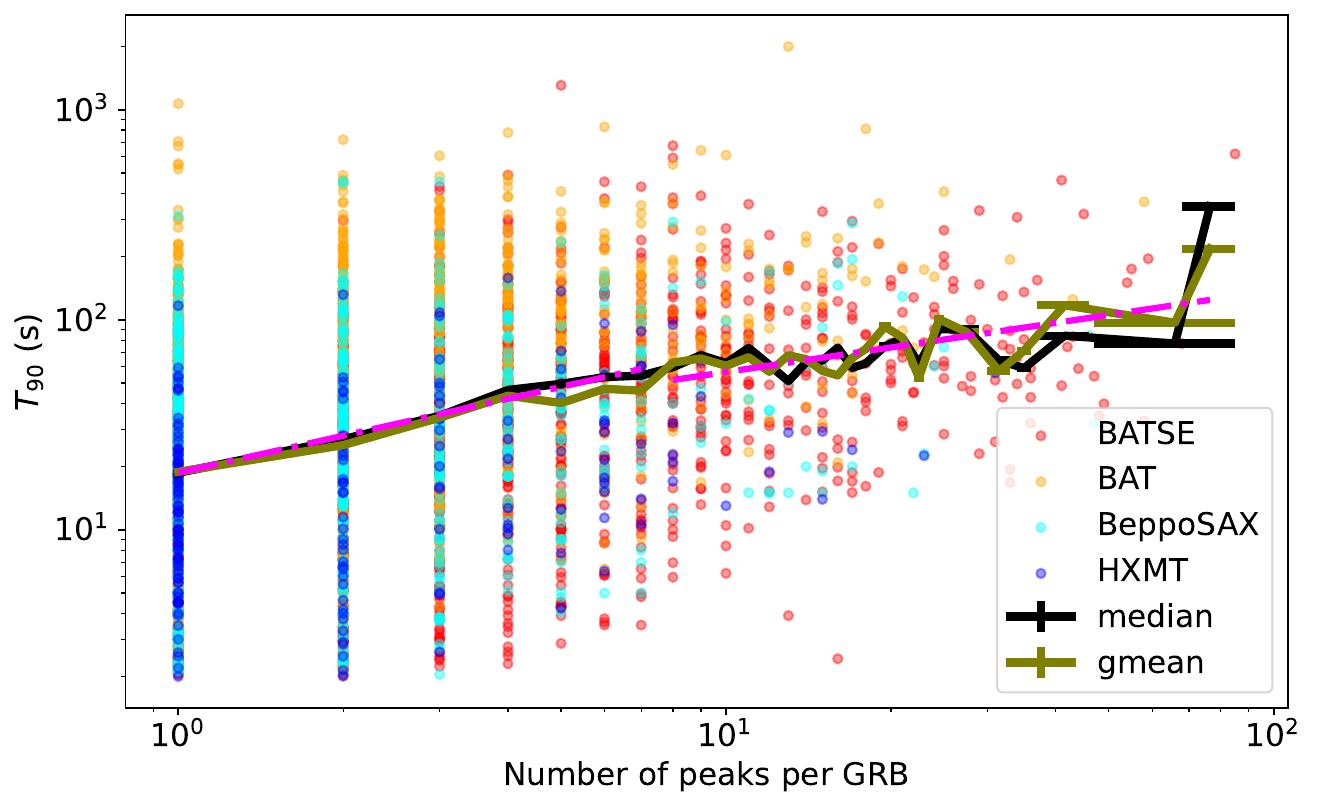}
    \caption{$T_{90}$ vs. number of peaks per GRBs for the different data sets. Also shown are the $T_{90}$ median (black) and the geometric mean (olive) as a function of number of peaks, grouped with at least 9 GRBs each. The dashed-dotted line (magenta) shows a broken power-law fit of the median behaviour.}
    \label{fig:T90vsNp}
\end{figure}
Similar results were obtained on a preliminary sample of 393 {\it Fermi} Gamma--ray Burst Monitor (GBM; \citealt{Meegan09}) GRBs, 361 of which are shared with the BAT sample (Maccary et al. in prep.): in particular, we found a peak-rich fraction $\bar{w}_2 = 0.19_{-0.09}^{+0.13}$, in line with the other four samples.

We investigated the relation between number of peaks per GRB $n$ and burst duration as expressed in terms of $T_{90}$, whose values were taken from the official corresponding catalogue papers. Figure~\ref{fig:T90vsNp} shows the results for all catalogues. We studied how the median and the logarithmic mean of $T_{90}$ of all catalogues merged together vary as a function of $n$. To avoid low-count fluctuations in the median and mean estimates, we grouped bins of $n$ so as to have at least 9 GRBs per class. The two quantities clearly track each other very closely (Fig.~\ref{fig:T90vsNp}). Interestingly, the dependence of $T_{90}$ on $n$ becomes shallower for $n\gtrsim n_2$: modelling the median value of $T_{90}$, $\tilde{T}_{90}$, with a power-law, $\tilde{T}_{90}/{\rm s}  \simeq \kappa\, n^a$, we found $a=0.58$ ($\kappa=18.8$) for $n<n_2$,
while $a=0.39$ ($\kappa=23.0$) for $n>n_2$ (dash-dotted magenta line in Fig.~\ref{fig:T90vsNp}). The fact that $a$ significantly differs from 1 suggests that the abundance of peaks within a GRB is not compatible with a common probability per unit time of emitting a peak, which would predict $a\sim 1$. Moreover, the dynamic range of $\tilde{T}_{90}$ (19~s for $n=1$, 57 s for $50<n<85$) is significantly narrower than the scatter of $T_{90}$ for any $n$: this indicates that $T_{90}$ is not a main driver of $n$. In addition, the fact that the dependence of $\tilde{T}_{90}$ becomes even shallower for large values $n$, where most GRBs are contributed by the peak-rich component of Eq.~\ref{eq:exp2}, further supports the possible existence of two families of GRBs or, at least, two different behaviours of GRB engines.

\subsection{Instrumental selection effects}
\label{sec:sel_eff}
The number of peaks clearly depends on the detector sensitivity and on the threshold on S/N. Secondly, it also depends on the energy passband, given that harder channels exhibit narrower and spikier temporal structures than softer channels.
This is clearly shown by the higher number of peaks in the BATSE sample compared with the other samples and is due to the better sensitivity of the former instrument.

We investigated the impact of S/N by repeating the analysis assuming two additional threshold values: S/N$>7$ and S/N$>9$.
The total number of identified peaks decreased to a fraction between 70 and 85\% for the different catalogues, when moving from S/N$>5$ to S/N$>9$. Only the average number of peaks per GRB of the BATSE sample decreased from $5.6$ to $4.9$, whereas for all the other samples the same quantity remained in the range $2.8$--$3.0$ peaks per GRB: the reason is that for the other three samples the number of GRBs featuring at least one significant peak decreased proportionally, due to lower sensitivity with respect to BATSE. Because of this, the weighted average number of peaks per GRB of both peak-poor and peak-rich components did not change significantly: $\langle n^{(1)}\rangle$ passed from $2.1\pm 0.1$ (S/N$>5$) to $2.0\pm 0.1$ (S/N$>9$). $\langle n^{(2)}\rangle$ changed from $8.3\pm 1.0$ (S/N$>5$) to $8.1\pm 1.0$ (S/N$>9$). Finally, the fraction of peak-rich GRBs moved from $0.19\pm 0.03$ (S/N$>5$) to $0.22\pm 0.03$ (S/N$>9$). Overall, while the number of peaks per GRB is inevitably expected to decrease by keeping raising the S/N threshold (although in a limited way, as we found passing from 5 to 9), the fractions of peak-poor and peak-rich GRBs do not seem to depend sensitively on the S/N threshold.  

\begin{figure}
   \includegraphics[width=\columnwidth]{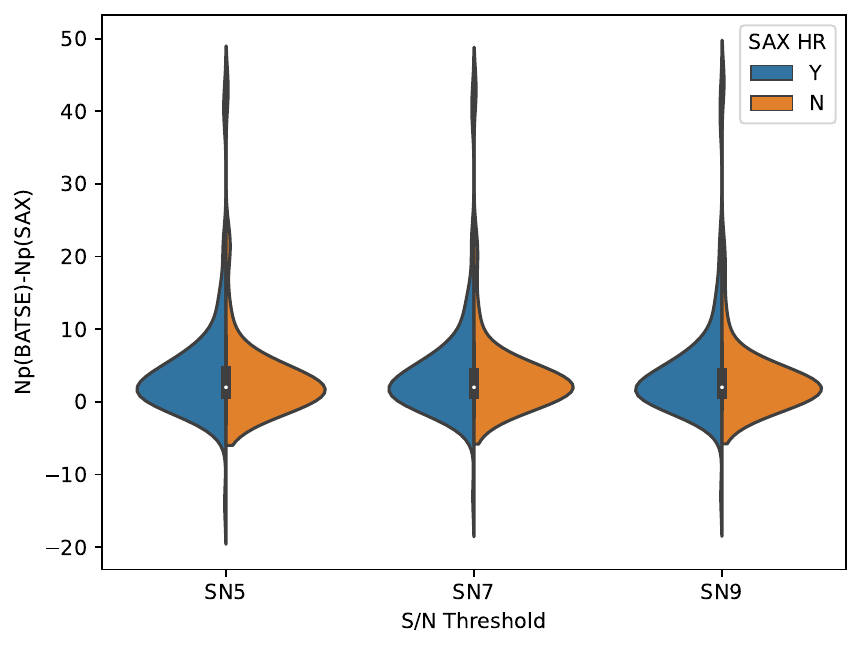}
    \caption{Distributions of $\Delta n$, defined as the difference between the number of peaks detected with BATSE and the one detected with {\it BeppoSAX} for a common sample of 160 GRBs, for three different thresholds on S/N. Blue and orange show the subsamples of GRBs with/without {\it BeppoSAX} high-resolution data, respectively.}
    \label{fig:BATSEvsSAX_violin}
\end{figure}

We identified 160 GRBs in common between the BATSE and the {\it BeppoSAX} samples. For each of these GRBs we calculated the difference between the number of peaks detected from the two profiles, $\Delta n = n^{\rm (BATSE)}-n^{\rm (SAX)}$. We divided this common sample into two subsamples: the GRBs for which there are high-resolution (HR; $62.5$-ms) data, and the GRBs with only 1-s data from {\it BeppoSAX}. Figure~\ref{fig:BATSEvsSAX_violin} shows the distributions of $\Delta n$ for different S/N thresholds.
First, the distributions do not depend significantly on the S/N threshold. Secondly, on average the presence of HR data from {\it BeppoSAX} does not impact appreciably the number of detected peaks. Thirdly, the median values in all cases is $+2$, which reflects the better sensitivity of BATSE. In addition, $\sim 60$\% ($\sim 70$\%) of all common GRBs have $|\Delta n|\le 2$ ($|\Delta n|\le 3$). A discrepancy of $\sim 2$--$3$ in estimating $n$ is small compared with $\langle n^{(2)}\rangle - \langle n^{(1)}\rangle \simeq 6.2\pm 1.0$. 
We derived the analogous distribution for the preliminary common sample of 361 GRBs shared by BAT and {\it Fermi}/GBM, confirming that for the majority of bursts the difference in the number of peaks is small: 86\% and 92\% GRBs with $|\Delta n|\le 2$ and $|\Delta n|\le 3$, respectively (Maccary et al. in prep.).
Consequently, for most GRBs the probability of belonging to either the peak-poor or the peak-rich family does not crucially depend on the detector used, at least as long as these two catalogues are concerned.

A more thorough characterisation through mapping with other GRB observed properties goes beyond the scope of the present analysis and would deserve a separate investigation.

\subsection{Number of peaks vs. peak luminosity}
\label{sec:Np_Lpiso}
According to the variability-luminosity relation, more luminous GRBs are, on average, more variable. We therefore studied the relation between number of peaks and peak luminosity.
To this aim, we selected 17 GRBs from the {\it BeppoSAX} and 131 from the BAT samples with spectroscopically measured redshift and for which the isotropic-equivalent peak luminosity $L_{\rm p,iso}$ in the comoving-frame $1$--$10^4$~keV passband had been estimated with broadband experiments, such as {\it BeppoSAX} and Konus/{\it WIND}. These values were taken from \citet{Ghirlanda05,Amati08,Yonetoku10,Dichiara16,KWGRBcat17,KWGRBcat21}.
\begin{figure}
   \includegraphics[width=\columnwidth]{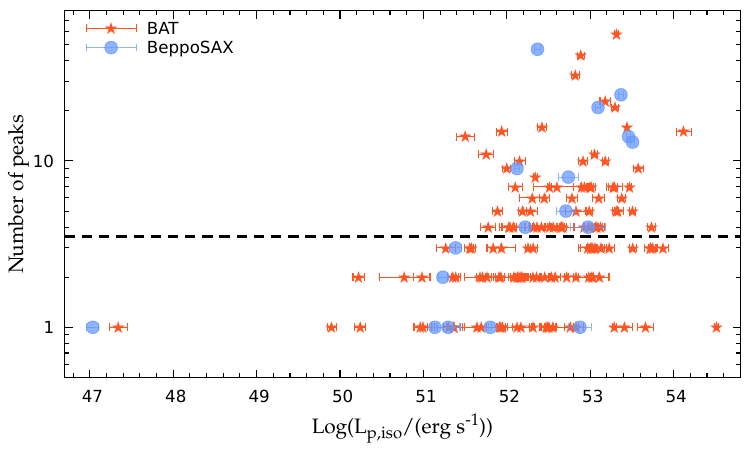}
    \caption{Number of peaks vs. isotropic-equivalent peak luminosity $L_{\rm p,iso}$ for a sample of 131 (17) GRBs with known $z$ from the BAT ({\it BeppoSAX}) samples. The dashed line marks the median number of peaks of the whole sample.}
    \label{fig:Np_vs_Lpiso}
\end{figure}
Figure~\ref{fig:Np_vs_Lpiso} shows the result. Non-parametric Spearman's and Kendall's correlation tests over the joint sample of 148~GRBs yielded a null hypothesis probability of no correlation as small as $1.1\times10^{-7}$ and $2.0\times10^{-7}$, respectively.
Hence, even though it is highly dispersed, the correlation is statistically significant, with peak-rich GRBs being, on average, more luminous.

As a further check, we also split the sample based on the median number of peaks of $n=3$ of the joint sample, ending up with 79 (69) GRBs with $n\le 3$ ($n>3$) peaks. The two groups can be roughly considered as peak-poor and peak-rich samples. Median peak luminosities are $1.5\times10^{52}$ and  $6.8\times10^{52}$~erg~s$^{-1}$ for the peak-poor and peak-rich samples, respectively. The probability of a common parent population for the two distributions of $L_{\rm p,iso}$ is $9\times10^{-5}$ and $<10^{-3}$ according to the Kolmogorov-Smirnov and Anderson-Darling tests,\footnote{These tests were done using {\tt scipy.stats.ks\_2samp} and {\tt scipy.stats.anderson\_ksamp}, respectively.} respectively. In conclusion, robust evidence is found that peak-rich GRBs are, on average, more luminous than peak-poor ones, although the two peak luminosity distributions are strongly overlapped.

\section{Discussion and conclusions}
\label{sec:disc_conc}
\begin{figure*}[!h]
   \includegraphics[width=\columnwidth]{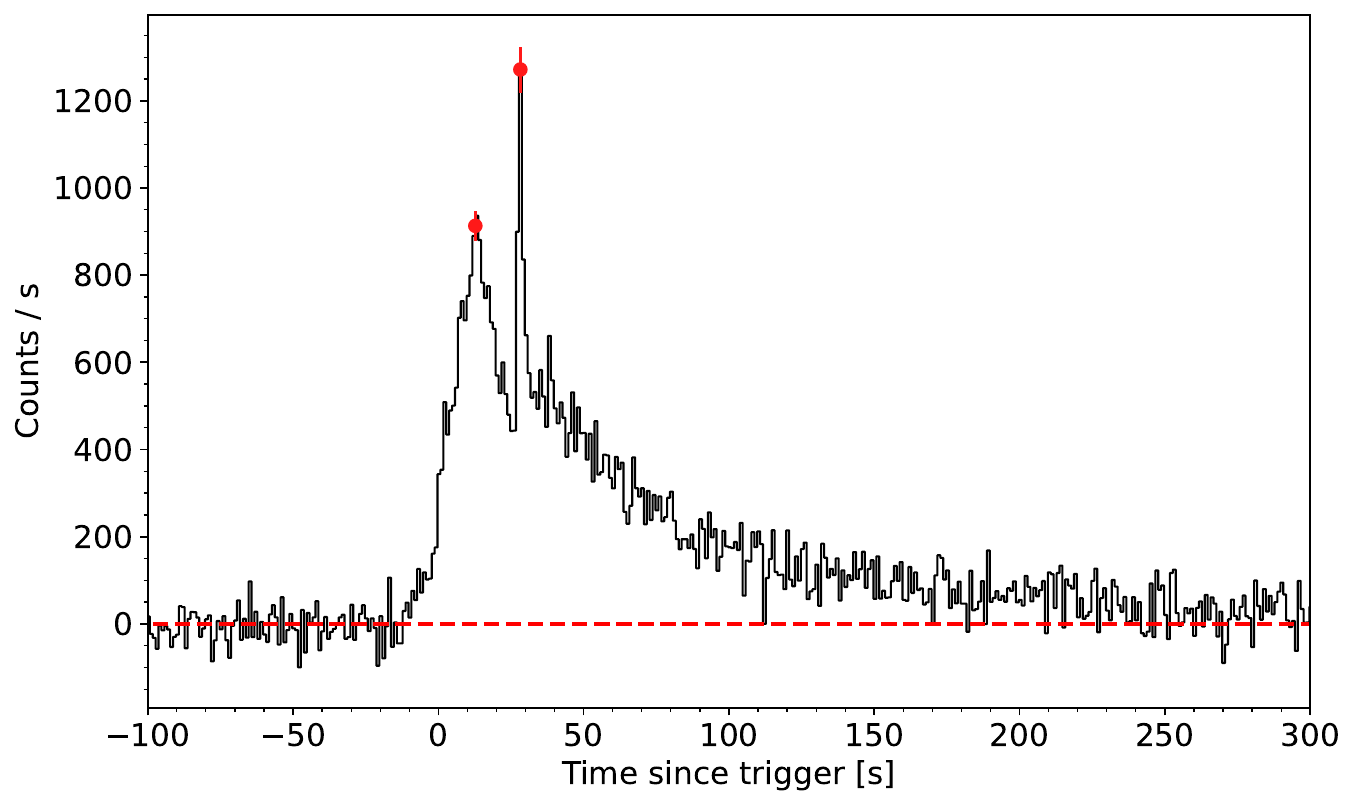}
   \includegraphics[width=\columnwidth]{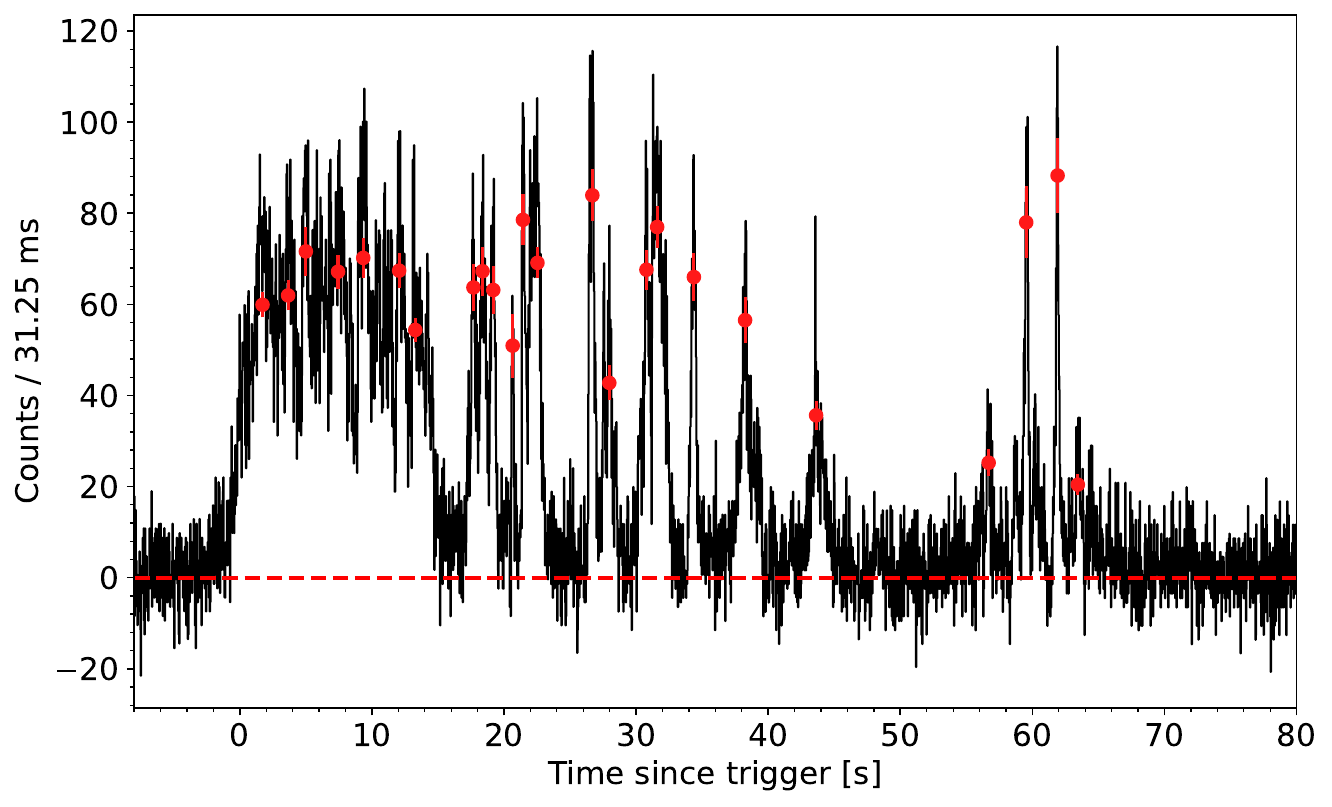}
    \caption{Examples of two GRBs having only 2 ({\it left}, 981203A) and 24 peaks ({\it right}, 010412), respectively, as observed in the 40--700~keV passband by {\it BeppoSAX}/GRBM. The red points show the identified peaks, while dashed lines show the interpolated background. The fluence and peak flux of the two GRBs are very similar \citep{Frontera09}.}
    \label{fig:LCexamples}
\end{figure*}

The analysis of a former {\it Swift}/BAT sample of GRB LCs showed that the power density spectra of peak-rich GRBs tend to be preferentially described by a shallow PL, which is the result of the superposition of peaks covering a broad range of timescales, such that no dominant timescale emerges (see Figure~7 of \citealt{Guidorzi16}). In particular, peak-rich GRBs tend to have relatively more temporal power on shorter ($\lesssim 1$~s) timescales than peak-poor GRBs due to the presence of a number of narrow peaks which are mostly absent in peak-poor GRBs. An illustrative example of this difference is shown in Figure~\ref{fig:LCexamples}: two bursts from the {\it BeppoSAX}/GRBM catalogue having comparable fluence and peak flux, one with just 2 peaks and a long smooth decaying tail ($T_{90}\sim 140$~s), and the other featuring 24 peaks within a $T_{90}=60$~s, most of which have subsecond durations.

In light of the results here reported on the possible existence of two distinct dynamics, as revealed by the distribution of number of peaks per GRB, the emerging picture is suggestive of two kinds of regimes through which Type-II GRB engines release their energy: (i) one emitting a few ($\sim 2.1$ on average), typically broad peaks ($\gtrsim 1$~s), makes up $\sim 80$\% of the observed population of GRBs of past and present catalogues; (ii) the other one emitting several peaks ($\sim 8.3$ on average), covering a broader range of timescales, including subsecond ones, which makes up the remaining 20\% of the observed population. The coexistence of two kinds of timescales in some GRBs, the so-called slow and fast variability components, was already suggested in the literature \citep{Vetere06,Margutti09,Gao12,Guidorzi16}: our results suggest a further characterisation, which points to the existence of two distinct ways GRB engines may work.

One can look at the class of peak-rich GRBs as being characterised by the presence of a fast component that manifests itself through a number of narrow peaks, which are instead mostly missing in the other class of peak-poor GRBs. 
Peak richness is also preferentially accompanied, on average, by shorter MVT and higher peak luminosity, although the latter alone cannot be used to discriminate between the two classes, given the strong overlap shown in Fig.~\ref{fig:Np_vs_Lpiso}.
The origin of the fast component in the literature has been investigated within the context of different progenitor models and can be summarised in two main groups, depending on whether the fast component (a) reflects the inner engine activity or (b) is mainly driven by other factors. Within the context of the internal shock (IS) model \citep{Kobayashi97,Daigne98,Maxham09}, the number of peaks and the sequence of time intervals simply reflect the emission time history of the inner engine. Unlike the slow one, the fast component imprinted by the inner engine would not be quenched, as a purely hydrodynamic jet propagates through the stellar envelope \citep{Morsony10}. When the possibility of a magnetised outflow is considered, a weakly magnetised and intermittent jet, which is less subjected to mixing with the surrounding progenitor gas than a pure hydrodynamic jet, appears more promising to explain the observed variability, since the predicted GRB emission would be less inhibited \citep{Gottlieb20a,Gottlieb21b}.

Within the (a) scenario, our results suggest that either inner engines of peak-poor GRBs are quieter, especially at short timescales, or their progenitor structure is able to suppress the subsecond variability. This could be due to the different degree of magnetisation of the jet, which can affect the mixing and consequently the jet baryon load \citep{Gottlieb20b}. There are a number of interpretations of how a GRB engine like an accreting black hole (BH) could explain the fast component: for example, through fluctuations of the viscosity parameter and how they consequently propagate through the disk in a similar way to what is believed in the case of BH binaries and active galactic nuclei \citep{Lin16}. Alternatively, they could be caused by the fastest modes of magneto-rotational instabilities that drive and launch a magnetised jet through the Blandford–Znajek process \citep{Janiuk21}. Within these contexts, peak-poor GRBs could be indicative of a regime in which either viscosity fluctuations or fast instabilities are strongly inhibited.

Within the alternative (b) scenario, the fast component that characterises peak-rich GRBs originates at further distance from the inner engine, up to the GRB emission site. In the Internal-Collision induced MAgnetic Reconnection and Turbulence (ICMART) model \citep{ICMART}, a wind of moderately magnetised shells (with magnetisation $1\lesssim\sigma\lesssim 100$ at the GRB site) collide and produce several runaway cascades of magnetic reconnection events. The fast component would be the result of turbulent, Doppler-boosted emission of mini-jets within the comoving frame of the colliding shells, caused by particles accelerated by magnetic reconnections. Simulations carried out by \citet{Zhang14b} show that spikier GRB LCs with many distinguishable peaks are, in particular, favoured by the following factors: (1) large emission site ($R\gtrsim 10^{15}$~cm) from the progenitor giving a longer curvature decay tail, which thus does not wash out the rapid variability due to the mini-jets; (2) a more magnetised outflow, which corresponds to a higher contrast between the Lorentz factors of the mini-jets and that of the corresponding shell. Our results could suggest that the turbulent magnetic reconnection regime either sets in efficiently (peak-rich GRB), or it does not (peak-poor GRB).

Ultimately, the possibility remains that the two classes reflect two different kinds of inner engines, such as ms-magnetars and accreting BHs, for which other potentially discriminating signatures in the prompt emission are yet to be identified.

A final note concerns the long-lasting GRBs that are Type-I candidates, such as 060614, 211211A, and 230307A \citep{Gehrels06,Yang22,Troja22,Gompertz23,Dichiara23,Levan23}, which were ignored in the present analysis: all their LCs exhibit numerous peaks (several dozens). While peak-richness alone cannot be a distinctive feature of this kind of merger candidates, given that there are several Type-II GRBs with an associated SN that are also peak-rich (e.g., 080319B, \citealt{Racusin08}), nevertheless it could be interpreted as a clue on the nature of the inner engine that powers these mergers.

\begin{acknowledgements}
We acknowledge the anonymous reviewer for a constructive report which improved the manuscript. C.G. acknowledges the Dept. of Physics and Earth Science of the University of Ferrara for the financial support through the ``FIRD 2022'' grant. M.B. is supported by the Dept. of Physics and Earth Science at the University of Ferrara through the FIRD 2022 grant. A.T. acknowledges financial support from ``ASI-INAF Accordo Attuativo HERMES Pathinder operazioni n. 2022-25-HH.0''.
\end{acknowledgements}


\end{document}